# Adaptive De-noising of Photoacoustic Signal and Image based on Modified Kalman Filter

Tianqu Hu#, Zihao Huang#, Peng Ge, Feng Gao, and Fei Gao*

Tianqu Hu, Zihao Huang, Peng Ge, Feng Gao and Fei Gao are with the School of Information Science and Technology, ShanghaiTech University, Shanghai 201210, China (# are co-first authors, *corresponding author: gaofei@shanghaitech.edu.cn)

*Abstract*—As a burgeoning medical imaging method based on hybrid fusion of light and ultrasound, photoacoustic imaging (PAI) has demonstrated high potential in various biomedical applications recently, especially in revealing the functional and molecular information to improve diagnostic accuracy. However, stemming from weak amplitude and unavoidable random noise, caused by limited laser power and severe attenuation in deep tissue imaging, PA signals are usually of low signal-to-noise ratio (SNR), and reconstructed PA images are of low quality. Despite that conventional Kalman Filter (KF) can remove Gaussian noise in time domain, it lacks adaptability in real-time estimating condition due to its fixed model. Moreover, KF-based de-noising algorithm has not been applied in PAI before. In this paper, we propose an adaptive Modified Kalman Filter (MKF) targeted at PAI de-noising by tuning system noise matrix Q and measurement noise matrix R in the conventional KF model. Additionally, in order to compensate the signal skewing caused by KF, we cascade the backward part of Rauch-Tung-Striebel smoother (BRTS), which also utilizes the newly determined Q. Finally, as a supplement, we add a commonly used differential filter to remove in-band reflection artifacts. Experimental results using phantom and *ex vivo* colorectal tissue are provided to prove the validity of the algorithm.

## I. INTRODUCTION

With the ability of noninvasively realizing structural and functional imaging for biological tissue, photoacoustic imaging (PAI), a burgeoning hybrid imaging modality, has shown significant potential in clinical applications [1-6]. However, stemming from PA signal's weak amplitude and strong random noise from external instrument and environment, PA signals and images obtained originally show low signal-to-noise ratio (SNR).

To overcome this challenge, there already exist numerous methods to eliminate white noise. The most common one is data averaging, which is already used in PAI. However, it requires extra room for data storage and is highly time-demanding [7]. Another strategy is classical Kalman Filtering (KF) realized in time domain. It predicts value from the past data and modifies the predicted result by the observed data. Therefore, the effectiveness of conventional KF relies on the correct definition of two key parameters, system noise matrix (Q) and measurement noise matrix (R), which are generally defined according to priori knowledge, and remain fixed throughout the processing run [8]. However, prior statistics are usually difficult to be exactly obtained in practical situation.

Conventional KF lacks adaptability since that model with constant parameters can hardly correspond to real-time variation and any external disturbance, which may easily cause instability and inaccuracy of the estimation system [9-10].

For improvement, several adaptive Kalman Filter strategies have been explored by researchers, two of which are Multiple Model Adaptive Estimation (MMAE) [11-12] and Innovation-based Adaptive Estimation (IAE) [10]. In the former one, a bank of Kalman filters runs in parallel under different models for the statistical filter information matrices. The modification of Q in the proposed algorithm refers to the principle of MMAE. While in the latter method, Q and R are adapted as measurements evolving with time, but it may increase storage burden due to that the innovation vector or residual vectors must be used for *m* epoch.

Neither conventional KF nor adaptive KF has been applied in PAI de-noising before. In this paper, we introduce an adaptive Modified Kalman Filter (MKF) to process PA signals. System noise matrix Q and measurement noise matrix R are adjusted for improving the stability and adaptability of KF. Experimental results demonstrate significant improvement of PA image quality.

## II. METHODS

In this paper, every single A-scan signal is processed point by point. Modified Kalman Filter (MKF) is the core of this proposed de-noising algorithm. In order to compensate signal skewing caused by MKF, a backward Rauch-Tung-Striebel smoother (BRTS) is then cascaded. As a supplement, a common used differential filter is added to wipe out in-band noise (noise with the same spectrum as PA signals) through subtraction and assist in PA image quality improvement.

### A. Modified Kalman Filter

*1) Theory of MKF*

The theoretical de-noising principle of MKF is the same as conventional Kalman Filter [14], which is shown below.

$$P_k^- = \mathrm{F}_{k-1}P_{k-1}^+\mathrm{F}_{k-1}^T + Q_{k-1}$$

$$K_k = P_k^- H_k^T (H_k P_k^- H_k^T + R_k)^{-1}$$

$$\hat{x}_k^- = F_{k-1}\hat{x}_{k-1}^+ + \mathrm{G}_{k-1}\mathrm{u}_{k-1} \tag{1}$$

$$\hat{x}_k^+ = \hat{x}_k^- + K_k(y_k - H_k\hat{x}_k^-)$$

$$P_k^+ = (I - K_kH_k)P_k^-$$

where $\hat{x}_{k-1}^+$ is the refined estimate after processing the measurement at time (k-1); $P_{k-1}^+$ is the covariance of $\hat{x}_{k-1}^+$; $\hat{x}_k^-$ is the estimate before processing the measurement at time k; $P_k^-$ is the covariance of $\hat{x}_k^-$; $\hat{x}_k^+$ is the refined estimate after processing the measurement at time k; $P_k^+$ is the covariance of $\hat{x}_k^+$; $F_{k-1}$ is the state transition matrix; $F_{k-1}^T$ is the transpose of $F_{k-1}$; $G_{k-1}$ is the control matrix which decides how $u_{k-1}$ works on the present state; $u_{k-1}$ is the control quantity; $H_k$ is a transition matrix that mathematically connects measurement form and state form; $H_k^T$ is the transpose of $H_k$; $K_k$ is the Kalman filter gain, which represents that algorithm believes more in measurement or prediction; $Q_{k-1}$ is the prediction noise covariance matrix; $R_k$ is the measurement noise covariance matrix; $I$ is identity matrix.

According to Eq. (1), Q and R both affect the value of Kalman gain $K_k$, and $K_k$ has a direct impact on the state estimation result. Specifically, if the collected PA signals suffer great noise disturbance, it will cause a large R and a small $K_k$, which means that the algorithm depends less on measurement and more on model prediction. On the contrary, a larger $K_k$ leads to higher dependence on measurement result. However, as mentioned in Introduction, traditional KF lacks adaptability because its parameters are set constant, thus failing to follow real-time change.

Moreover, since every A-scan signal is processed point by point, all matrices in formulas are of one dimension and can be regarded as a number. Thus, Q and R become system noise number and measurement noise number respectively. Then we introduce optimization methods to modify Q and R.

### 2) Predetermine System Noise Number Q

By simulation attempts, we found that Kalman filter and BRTS with different Q values may cause amplitude attenuation on each A-scan PA signal to different extent. Therefore, in pursuit of avoiding amplitude distortion, we need to decide a suitable Q that the whole system can share. An alternative way referring to the principle of MMAE [11-12] is proposed.

We randomly choose a certain number of A-scan signals, and then parallelly run MKF with BRTS models with a certain range of Q values. After obtaining the group of 'Q's with the best SNR, we use their mean value as the final Q for the whole algorithm. The flow chart below demonstrates the process of adaptive predetermination of Q (Fig. 1).

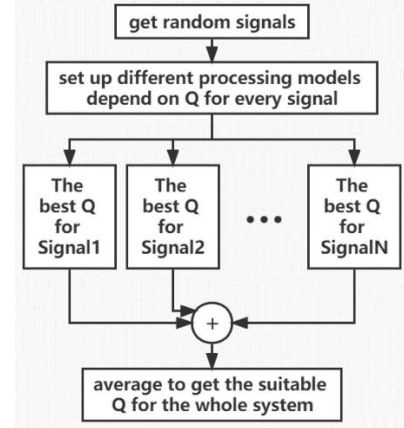

**Fig. 1.** The process of adaptively predetermining Q.

### 3) Determine Measurement Noise Number R

In this paper, measurement noise number R represents the noise power of collected signals. Since every A-scan PA signal is individually processed, R should be computed separately for each obtained A-scan signal. In detail, we take the beginning data points of the PA signals, from the first point to the one before effective PA pulse, as noise points (Fig. 2), which contain white noise information. All of these noise points are noted as Nd, and the noise power can be calculated by below equation:

$$R = \frac{Nd*Nd^T}{length\ of\ Nd} \qquad (2)$$

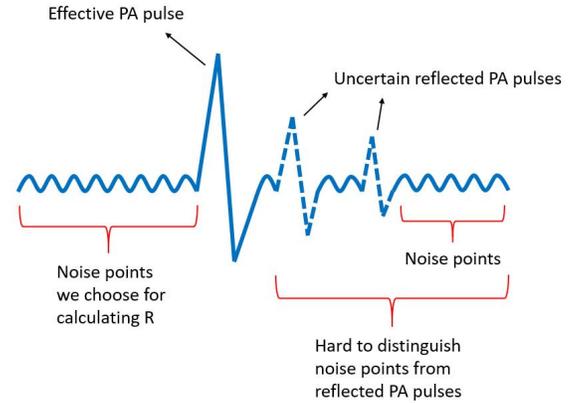

**Fig. 2.** The sketch map of one A-scan signal and the choice of noise points.

Due to various reflection intensity and reception angles of UT, the reflected PA pulses may present uncertain amplitudes and appear in different locations, displayed in dotted lines in Fig .2. As a result, it is hard to distinguish noise points from the uncertain reflected PA pulses.

Moreover, since that the repetition rate of laser system is 1 kHz, the sampling process of one single A-scan signal is completed almost instantaneously. During this transient period, the information contained in the noise points (excluding reflected PA pulses) remains consistent in the whole signal.

Therefore, we choose the beginning data points of an A-scan signal as noise points for calculating R in MKF.

### B. Cascaded backward RTS smoother (BRTS)

As a type of tracking algorithm, MKF may cause certain degree of phasor lagging, or called skewing, which may lead to depth misjudgment of the sample. Just contrary to the process sequence of MKF, BRTS [14] starts from the final point of an A-scan PA signal. It predicts the $N^{th}$ datapoint from the $(N+1)^{th}$ data point and corrects the results we obtained by MKF. It does not rely on measurement data, but still uses the predetermined Q while working. By combining the process results of MKF forward filtering ($\hat{x}_{kN}^+$) and BRTS ($\hat{x}_{bN}^+$), the final estimation result of time N can be obtained (Fig. 3). During this process, not only are signals smoothed, but the phasor skewing is corrected as well. Then the depth of tissue in region of interest (ROI) can be distinguished clearly.

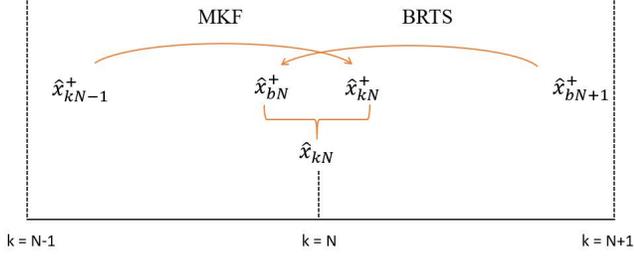

**Fig. 3.** The illustration of how MKF and BRTS cooperate.

### C. Supplemented Differential Filter

After removing Gaussian noise in PA signals, we append a differential filter as a supplement to treat in-band noise that cannot be eliminated by MKF. The strategy is illustrated in Fig. 4.

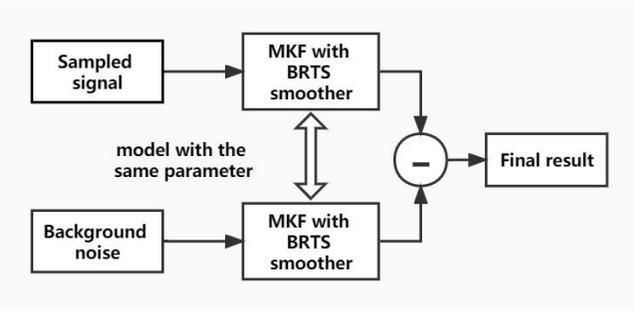

**Fig. 4.** The realization of differential filter.

We first collect background noise signals from phantoms without charcoal sticks and the carrier of colorectal tissues within a pre-set ROI. After that, we collect PA signals from the samples of phantoms with charcoal sticks and *ex vivo* colorectal tissues. For keeping consistency of PA signals and background noise while using differential filter, we process both of them by MKF with BRTS with the same model parameters. The final result is obtained by subtracting background signal from the corresponding PA signal.

## III. Experimental Setup And Results

### A. Experimental Setup: PAM system

The experimental setup to validate the effectiveness of the algorithm is photoacoustic microscopy (PAM) system, whose schematic is displayed in Fig. 5. The 532 nm wavelength light is first emitted by the pulsed laser. Objective lens and multiaxial displacement device are used to ensure the laser's entry efficiency into the fiber. In pursuit of minimizing the spread loss, we use two condenser lenses to make sure that the laser is focused exactly on the sample. The laser intensity is set at 25 dB while collecting PA signals. The center frequency of the ultrasound transducer is 2.5 MHz.

During experiments, PA signals are generated from the sample and collected by ultrasound transducer (UT). After amplificated by a low-noise amplifier, PA signals are digitized by an oscilloscope and then stored in the computer. By raster scanning, the reconstructed image has a total pixel number of 250×100×2048 in our experiment.

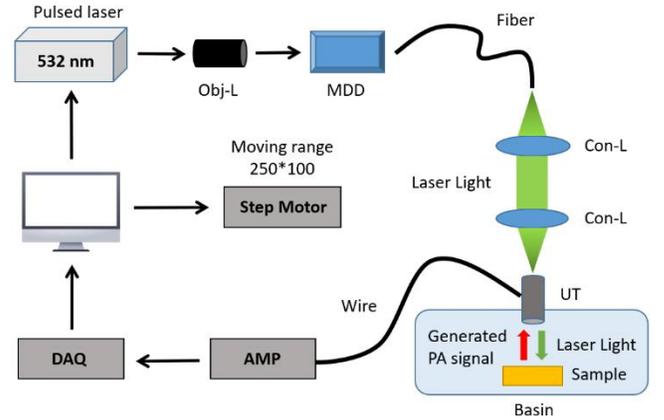

**Fig. 5.** The schematic of the PAM system. Obj-L: objective lens, Con-L: condenser lens, MDD: multiaxial displacement device, UT: ultrasound transducer, AMP: amplifier, DAQ: data acquisition.

### B. Experimental Results

In order to preliminarily verify the feasibility of our algorithm, we performed experiment in phantom first. We choose agar and water as the raw material of phantoms, since the absorption and scattering coefficients of agar are similar to that of human skin and fleshy tissue. Then we insert charcoal sticks into prepared phantoms and lay out disparate shapes. After obtaining a certain number of satisfying outcomes, we then turn to *ex vivo* colorectal tissues imaging experiments. We process raw data with traditional lowpass filter (LP) to make a comparison. Measurement results of both experiments are shown below.

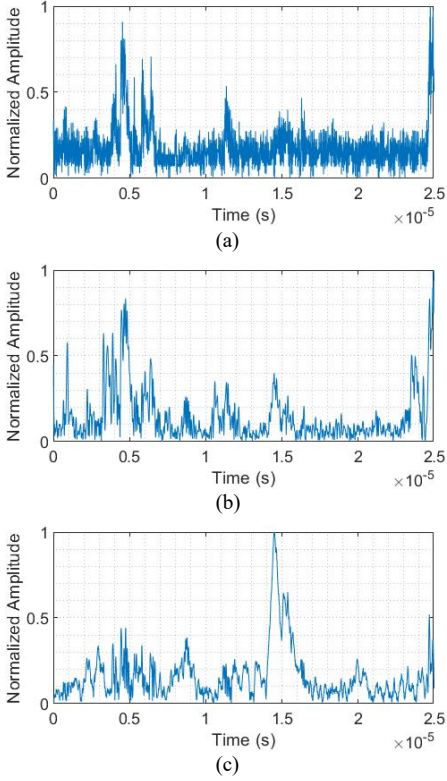

**Fig. 6.** (a) Hilbert Transform of collected original PA signal. (b) Hilbert Transform of signal processed by conventional lowpass filter and differential filter. (c) Hilbert Transform of signal processed by proposed algorithm.

The image reconstruction of PAM system applies Hilbert Transform (HT) to signals and then captures the highest amplitude to form PAM images. Therefore, we directly display the signals after HT.

According to the A-scan original signal in Fig. 6 (a), the effective PA pulse existing at $1.5*10^{-5}$ s is overshadowed by the abrupt high-amplitude noise interferences. Fig. 6 (b) shows that although low-pass filter with differential filter can remove noises to a certain extent, it is incapable of eliminating the prominent interferences. In contrast, after processed by our method, the effective PA signal stands out and all of the abrupt interferences are significantly suppressed in Fig. 6 (c).

To be more specific, we use PSNR to compare the denoising performance. The signal PSNR is calculated by Eq. (3) below, and the result comparison is illustrated in Fig. 7.

$$PSNR = 10 * \log\left(\frac{(maximum\ ROI\ signal)^2}{noise\ power\ outside\ ROI}\right) \quad (3)$$

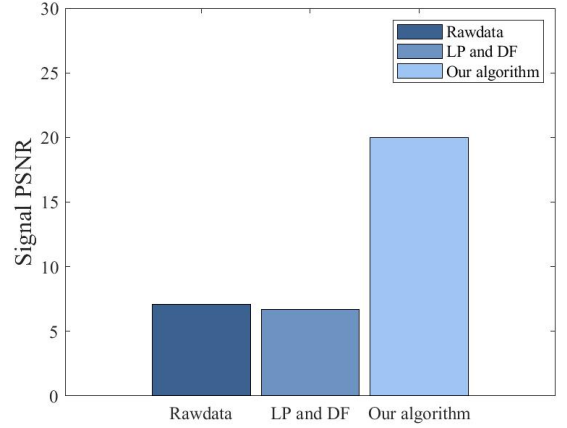

**Fig. 7.** Signal PSNR comparison.

Referring to the figure above, our algorithm increases the signal PSNR by 13.2 dB in average, comparing with the results obtained by the LP with DF method.

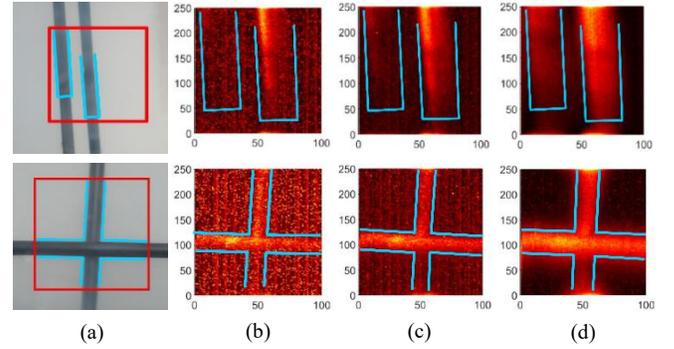

**Fig. 8.** (a) The photos of phantom, with ROI marked on it. (b) Original reconstructed images. (c) Reconstructed images after processing PA signals with LP and DF. (d) Reconstructed images after processing PA signals with our algorithm.

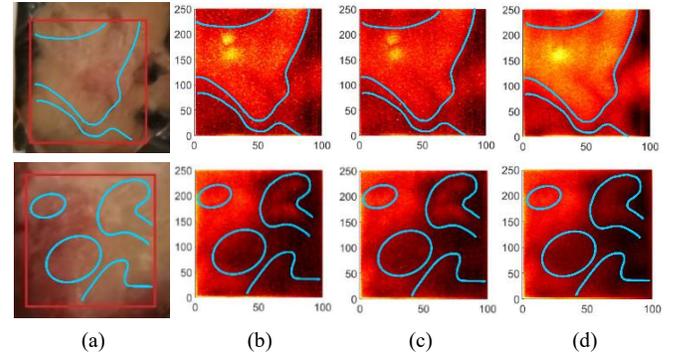

**Fig. 9.** (a) The photos of *ex vivo* colorectal tissues, with ROI marked on it. (b) Original reconstructed images. (c) Reconstructed images after processing PA signals with LP and DF. (d) Reconstructed images after processing PA signals with our algorithm.

Shown in Fig. 8 and Fig .9, two groups of image comparison are selected for phantom and *ex vivo* colorectal tissues respectively. The regions where more effective information becomes distinct after applying the proposed algorithm are outlined in blue. It is apparent that conventional LP with DF method causes little change to the image, and some details are even weakened. On the contrary, our method makes the internal curves of the image more obvious and the boundaries more distinct.

Combining with the signal level results, due to the existence of high-amplitude interferences, many abrupt bright spots are displayed, which make the images in Fig. 8 (b) (c) and Fig. 9 (b) (c) look coarse. After the distractions are eliminated by our method, PA signals can be recovered and background parts are much darker. As demonstrated in Fig. 8 (d) and Fig. 9 (d), not only the noisy points disappear, but patterns that should be highlighted turn to be brighter as well. Furthermore, because of the smoothing effect of MKF and BRTS, most strip-type system background noise (caused by the column-by-column sampling property) are observably suppressed.

## IV. CONCLUSION

In conclusion, we propose an adaptive de-noising algorithm based on Modified Kalman Filter with backward RTS smoother and supplemented by a differential filter. MKF aims to adaptively and accurately wipe out Gaussian noises in real time, while BRTS can solve the signal backward skewing caused by MKF. On the basis of the first two parts, a commonly used DF is then added to remove the in-band artifacts. For MKF, in order to avoid amplitude distortion, we predetermine a suitable system noise Q that the whole system can share. By means of appropriately choosing noise points of PA signals, measurement noise R are calculated for each single A-scan signal. By adjusting Q and R in the estimation model, MKF and RTS can adaptively follow the real-time change of noises. Phantom and ex vivo imaging experiments are separately performed to validate the effectiveness of this algorithm. Results prove that PA signals are perfectly restored and the quality of reconstructed PA images is significantly improved.

In addition, the structures of both MKF with BRTS and differential filter are concise. The concision makes it possible for us to realize the algorithm in hardware acceleration in the future. We also hope that more time-domain filtering methods, such as extended Kalman filter, particle filter, can be applied to achieve better filtering effect. In addition, we will further adapt the accelerated algorithm to more imaging scenarios, including PA computed tomography system, so as to expand the application scope and feasibility of the method.